# MT-lib: A Topology-aware Message Transfer Library for Graph500 on Supercomputers

## Xinbiao Gan

**Abstract**—We present MT-lib, an efficient message transfer library for messages gather and scatter in benchmarks like Graph500 for Supercomputers. Our library includes MST version as well as new-MST version. The MT-lib is deliberately kept light-weight, efficient and friendly interfaces for massive graph traverse. MST provides (1) a novel non-blocking communication scheme with sending and receiving messages asynchronously to overlap calculation and communication;(2) merging messages according to the target process for reducing communication overhead;(3) a new communication mode of gathering intra-group messages before forwarding between groups for reducing communication traffic. In MT-lib, there are (1) one-sided message; (2) two-sided messages; and (3) two-sided messages with buffer, in which dynamic buffer expansion is built for messages delivery. We experimented with MST and then testing Graph500 with MST on Tianhe supercomputers. Experimental results show high communication efficiency and high throughputs for both BFS and SSSP communication operations.

**Index Terms**—MST; Graph500; Tianhe supercomputer; Two-sided messages with buffer

———————————— ◆ ————————————

## 1 INTRODUCTION

THE development of high-performance supercomputing has always been a strategic goal for many countries [1,2]. Currently, exascale computing poses severe efficiency and stability challenges.

Large-scale graphs have applied to many seminars from the real world. Graph500 benchmark measures the data analysis performance based on graph computing and application, popularly testing on supercomputers. Tianhe-2 supercomputer has ranked top 1 in Top500 list for 6 times, but it failed to win Graph500 championship. In addition to poor locality and irregular memory access from BFS (Breadth-First Search), the key reason is that Matrix-2000+ and proprietary Interconnect built in Tianhe are not fully utilized for running Graph500. Consequently, NUDT (National University of Defense Technology) would committed to explore feathers from Graph500 and exploit powerful data processing potential to show the capacity of Tianhe Pre-exacale system as well as Tianhe exascale system.

Recent emergence of extremely large-scale graphs in various application fields including social networks, business intelligence, and public safety, requires fast and scalable graph analysis. With such high interest in analytics of large graphs, a new benchmark called the Graph500 was proposed in 2010[3]. Differently from the Top 500 used to supercomputers metric with FLOPS (Floating Point Per Second) for compute-intensive su-percomputing applications. The Graph500 benchmark instead measures data analytics performance on data-intensive applications, in particular those for graphs, with the metric TEPS (Traversed Edges Per Second).

Massive graph traverse including Graph500 has typical characteristics such as low parallel-ism, poor data locality and irregular memory access. Thus, we have conducted a considerable amount of research and works on Graph500 communications on distributed parallel systems for improving Graph500 performance [4-8].

In practice, optimizing benchmarks like Graph500 to run on supercomputers is a complicated process. Effective utilization of communication resources is vital for Graph500 performance and scaling efficiency. An efficient method is parallelizing Graph500 using MPI (Message Passing) Interface, a language-independent communication protocol that coordinates the computing tasks in parallel programs. But, MPI communication optimization requires a detailed understanding of the usage characteristics of applications on production supercomputing systems, especially for large-scale graph traverse such as BFS (Breadth-First Search) and SSSP (Single Source Shortest Path) [3]. Hence, a famous communication called AML (Active Message Library) is built in reference code for Graph500. Unfortunately, AML only support one-sided message without response, moreover, performance from AML does not have expecting behavior on supercomputers, especially for Tianhe Supercomputer.Accordingly, we present MST, an efficient message transfer library for messages gather and scatter for Graph500 on supercomputers.

The remainder of this paper is organized as follows. We will present related works including Graph500 and com-

————————————

• *Xinbiao gan is with the National University of Defense Technology, Changsha, China. E-mail: xinbiaogan@nudt.edu.cn*
• *S.B. Author Jr. is with the Department of Physics, Colorado State University, Fort Collins, CO 80523. E-mail: author@colostate.edu.*
• *T.C. Author is with the Electrical Engineering Department, University of Colorado, Boulder, CO 80309. On leave from the National Research Institute for Metals, Tsukuba, Japan E-mail: author@nrim.go.jp.*

***Please provide a complete mailing address for each author, as this is the address the 10 complimentary reprints of your paper will be sent***

*Please note that all acknowledgements should be placed at the end of the paper, before the bibliography (note that corresponding authorship is not noted in affiliation box, but in acknowledgement section).*

munication library in Section 2, and introduce the proprietary interconnect built in Tianhe supercomputers in section 3. Section 4 and Section 5 describe our methodology and implementations with evaluation, respectively. Section 6 provides a brief conclusion and discussion.

## 2 RELATED WORK

### 2.1 Graph500

The Graph500 is proposed to measure performance of a computer system for data-intensive applications that require an irregular memory access. Different from Top500, which is known as a list that ranks computers by running Linpack benchmark with GFLOPS (Giga FLOPS) for compute-intensive workloads. The Graph500 ranks computers by executing a set of data-intensive large-scale graph problems with GTEPS (Giga TEPS).

Until now, there are three ranks including BFS, SSSP (Single Source Shortest Path) and GreenGraph500 in graph500 benchmark [3], in which, BFS is the most famous and get widely attention. Graph500 performs BFS to a Kronecker graph modeling real-world networks from selected 64 roots randomly. Graph500 benchmark must perform the following steps according to specification and reference implementation. [3].

Step1(Edge Generation): This step produces the edge list using Kronecker graph generator according to recursively sub-divides adjacency matrix into 4 partitions A, B, C, D, then adding edges one at a time with partitions probabilistically A = 0.57, B = 0.19, C =0.19, D = 0.05 as Figure 1. This step is not timed for Graph500 performance.

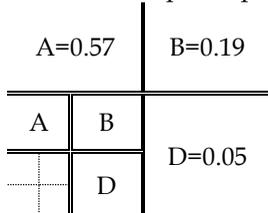

Figure 1. Recursively sub-divides adjacency matrix

Step2(Graph Construction): This step would construct a suitable data structure, such as CSR (Compressed Sparse Row) graph format for performing BFS from Step1(Edge Generation). In practice, this step is very crucial for performing BFS quickly, and graph construction is also not timed for Graph500 performance.

Step3(BFS searching): It is the key kernel to create a BFS tree and this step is the only one that should be timed for Graph500 performance.

Step4(Tree validation): Finally, Graph500 would verify the result of the BFS tree produced by Step3.

### 2.2 Communication Library

Communication performance is of paramount importance to high performance computing (HPC) applications. MPI (Message Passing Interface) is the predominant parallel programming model for supercomputers today, making it a key technology to be optimized so that scientific computing applications can take full advantage of the supercomputing system that they use. Optimization re-quires a detailed understanding of the usage characteristics of applications on production supercomputing systems. Unfortunately, the performance of MPI implementations on large-scale supercomputers is significantly impacted by factors including its inherent buffering, type checking, and other control overheads. Consequently, we have conducted a considerable amount of research and works on communication optimizations, especially for Graph500 with communication library running on Supercomputers.

In order to well understand MPI usage characteristics, Autoperf is created and some surprising insights we gathered from detailed analysis of the MPI usage logs, which reveal that large-scale applications running on supercomputers tend to use more communication and parallelism. While MPI library does not behavior well and performs discrepancy vary from supercomputers from different vendors. Accodingly, M. Blocksome et.al designed and implementated of a one-sided communication interface for the IBM Blue Gene/L supercomputer [9], which improved the maximum bandwidth by a factor of three. Furthermore, Sameer Kumar et.al presented several optimizations by extensively exploiting IBM Blue Gene/P interconnection networks and hardware features and enhancements to achieve near-peak performance across many collectives for MPI collective communication on Blue Gene/P [10]. Motivated for better support of task mapping for Blue Gene/L supercomputer, a topology mapping library is used in BG/L MPI library for improving communication performance and scalability of applications [11]. but, comprehensiveopology mapping library might benefit by providing scalable support of MPI virtual topology interface. Moreover, Kumar et.al presented PAMI (Parallel Active Message Interface) as Blue Gene/Q communicationlibrary solution to the many challenges with unprecedented massive parallelism and scale [12]. In order to optimize the performance of large-scale applications on supercomputers. IBM developed LAPI (Low- level Applications Programming Interface), which is a low-level, high-performance communication interface available on the IBM RS/6000 SP system [13]. It provides an activemessage-like interface along with remote memory copy and synchronization functionality. However, the limited set from LAPI does not compromise on functionality expected on a communication API, what's worse is that topology mapping library and LAPI is designed for IBM supercomputers, resulting in difficulties in adataption to applications running on general supercomputers, especially for Graph500 testing on other supercomputers.

Different from communication optimizations on IBM supercomputers, Naoyuki Shida et.al implemented a customized MPI library and low-level communication at tofu topology level based on open MPI for K supercomputer [14-15]. Similar with IBM supercomputers, above proposed MPI implementation is target at K supercomputer.

For insight into performance influence on Graph500 from MPI communication, Mingzhe Li et.al presented a detailed analysis of MPI Send/Recv and MPI-2 RMA based on Graph500 and exposed performance bottlenecks, furthermore, they proposed a scalable and high performance design of Graph500 using MPI-3 RMA to improve GTEPS (Giga TEPS) and win two times sppedup on TACC Stampede Cluster [16]. But, above analysis and efficient usage optimization on MPI are hardly established into a general Communication Library for Graph500 testing.

Fortunately, Anton Korzh from Graph500 executive committee committed AML (Active Messages Library) to opensource from Graph500 reference code [3]. AML is an SPMD (Single Program Multiple Data) communication library built on top of MPI3 intented to be used in fine grain applications like Graph500. AML would make user code clarity while delivering high performance, But, current version of AML support only one-sided message, which can not send a response from active message handler and there are no two-sided active messages, which would facilitate andbeneficial to hybrid graph traverse in Graph500.

Accordingly, we present MST (MeSage Tansfer) communication, an efficient message transfer library for messages gather and scatter in benchmarks like Graph500 for Supercomputers, especially for Graph500 testing on Tianhe supercomputers.

## 3 PROPRIETARY INTERCONNECT

Interconnection topologies play an important role in the supercomputers system. Currently, massively parallel computer systems have become popular [17-18], where the interconnection networks consist of a lot of processing cores. For example, Tianhe Pre-exascale system has 96608 processors located in 8 racks and AI testing platform in Tianhe scales up to 65K processors. Thus, the interconnection topologies in these systems have become a more critical factor for massive applications including benchmarks than the computing and memory subsystem. Communication hops, which largely depends on the interconnection topology, are of great concern in these supercomputer systems and has determined the behavior of Graph500, as well as Top500 with the growth of system sizes and shrink of clock cycles.

Accordingly, Tianhe Pre-exascale system and AI testing platform in Tianhe have adopted a proprietary interconnect network [2]. The network logic is developed and integrated into the network interface chip (named HFI-E) and the network router chip (named HFR-E). Both chips implement efficient mechanisms to achieve high-performance communication with regard to bandwidth, latency, reliability, and stability. The link rate is upgraded to 25 Gbps from 14 Gbps of the TianHe-2A supercomputer system.

HFI-E provides the software-hardware interface for accessing the high-performance network, implementing the proprietary MP/RDMA (Mini Packet/Remote Direct Memory Access) communication and collective offload mechanism. HFI-E contains a 16-lane PCIe 3.0 interface and connects with interconnect fabric via network ports.

HFR-E contains 24 network ports. Each port has an eight-lane 25 SerDes Gbps, with 200 Gbps unidirectional bandwidth. The throughput of a single HFR-E chip is up to 9.6 Tbps, and the Message Passing Interface latency is 1.1 us per one hop. HFR-E also adopts FC-PBGA (Flip Chip-Plastic Ball Grid Array) packaging technology and supplies 2816 pins.

The interconnection system network adopts a two-dimensional tree network topology on the basis of opto-electronic hybrid interconnection, as shown in Figure 2. The compute frames are connected by communication frames using active optical cables with a two-dimensional tree network topology. Such a topology is more advanced than n-D-Torus topology. The links between the adjacent nodes on each dimension are replaced with tree switches.

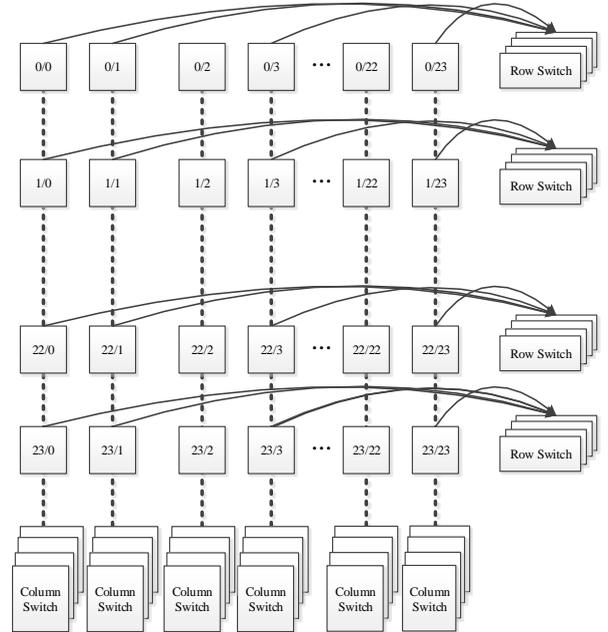

Figure 2. Network topology for Tianhe systems

Based on proprietary interconnect built in Tianhe families, we always prefer an efficient message transfer library for testing benchmarks like Graph500 as well as TOP500.

## 4 METHODOLOGY

AML is high performace communication library for graph computing, especially for Graph500 testing, but AML has only supported one-sided message. Furthermore, one-sided message from AML does not behavior well as expecting performance, especially for Graph500 with AML running on Tianhe supercomputers. Consequently, we redesign and rewrite MST based on AML according to proprietary interconnect built in Tianhe supercomputers. Different from AML, MST not only supports one-sided message but also holds two-sided messages. More importantly, two-sided messages with dynamic buffer expansion is also buit in MST.

### 4.1 One-sided Message

In Graph500 reference, AML has supported one-sided message, but it does not behavior well on Tianhe supercomputers. So we rewrite one-sided message according to proprietary interconnect built in Tianhe supercomputers.

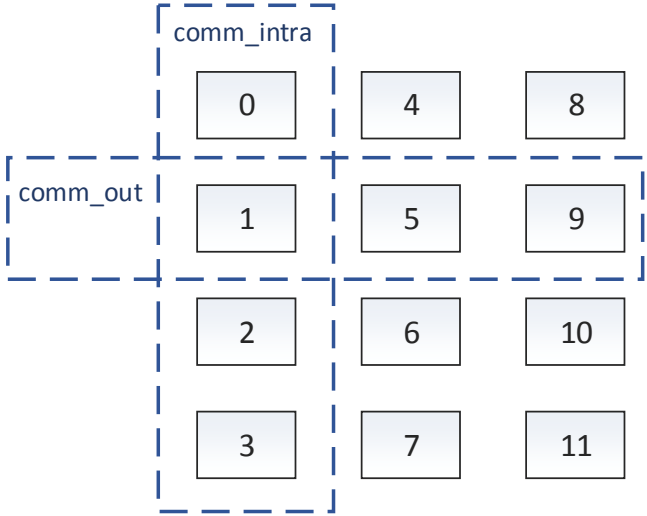

Figure 3. Communication domain division in AML

In AML, the global communication would be refined into two sub-communications including comm-intra and comm again. The processes with the same IP are painted with the same color, and then processes in the global communication domain are divided into a communication domain comm-intra according to this color. Moreover, processes with the same local process rank in each comm-intra are divided into a communication domain comm-inter, as illustrated in Figure 3. If all processes are regarded as a matrix, the processes on comm-intra form a column, and processes with the same local rank on each comm form a row. In the com, the group id of each process is its internal ID in the comm communication domain.

According to framework from communication domain division, AML drafts principle of one-sided message, in which message firstly transfer across comm sub-communications domain, and then forwarding in comm-intra sub-communications domain. Taking an example, regarding rank_id as process with rank=4, so rank_4 send message to rank_2 with color brown; and rank_1 transfer message to rank_11 painted color green, as shown in Figure 4.

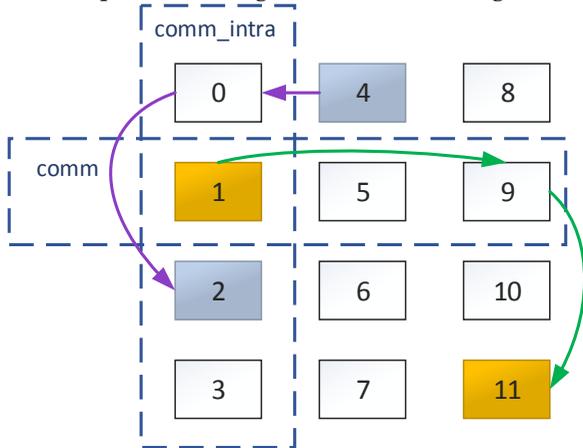

Figure 4. One-sided message in AML

As illustrated in Figure 4, message from rank_4 to rank_2 should firstly be sent to rank_0 across comm_intra sub-communications domain or in comm sub-communications domain, then rank_0 would forward message to destination rank_2 in comm_intra sub-communications domain, and so on, message from rank_1 to rank_11 should firstly be sent to rank_9 across comm_intra or in comm, then rank_9 would forward message to destination rank_11 in comm_intra sub-communications domain. In above message transfer flow from one-sided message, message firstly should be transfered across comm_intra, secondly would be forwarded in comm_intra, in other words, inter-node communication comm must before intra-node communication comm_intra. Obviously, sending and receiving message in comm_tra is superior to transferring message in comm. one-sided message in AML is a surprising mode that rejects what is close and seeks what is far. Accordingly, the MST is proposed, in which one-sided message mode is opposite to message transferring flow for AML.

Different from AML, in MST messages firstly should gather in comm_intra sub-communications domain, and then forward to destination across comm_intra or in comm_inter that is similar to comm in AML, as demonstrated in Figure 5.

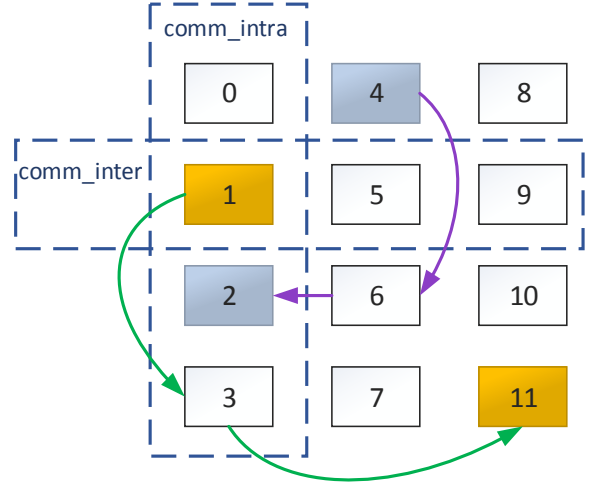

Figure 5. One-sided message in MST

Similarly, marking rank_id as process with rank=4, in Figure 5, there are two messages from rank_4 to rank_2 and from rank_1 to rank_11, respectively in Figure 5. message from rank_4 to rank_2 is advisable to firstly send message to rank_6 in same comm_intra, then rank_6 would forward message to rank_2 insteasd of opposite message flow in AML, and so on, message from ank_1 to rank_11 should firstly send to rank_3 in same comm_intra, and then rank_3 would forward message to rank_11.

Comparing AML and MST, it is easy to conclude that the main difference between MST and AML is message transferring flow, in which sending message across comm firsly then forwarding message in comm_intra from AML, while in MST sending message in comm_intra before forwarding message in comm_inter. Hence, both theoretical design an pratical work, tthe performace of MST is rather better than that of AML.

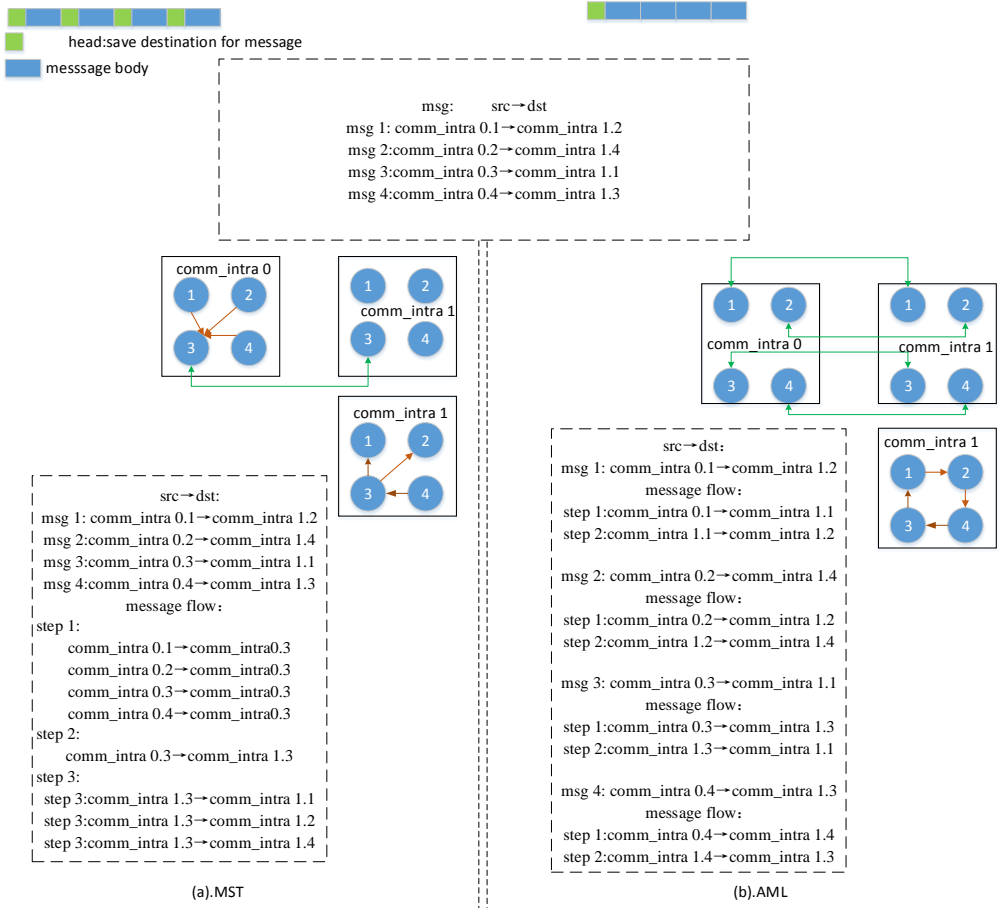

Figure 6. difference on message transferring policy between MST and AML

Theoretically, given the networking hops for message exchange in comm_tra is hops_intra, networking hops for message exchange across comm_tra is hops_inter, $s$ denotes the number of transferred messages. The accumulative hops $aml\_hops$ for one-sided message from AML and MST are taged as $AML\_hops$ and MST-hops listed in equation (1) and equation (2) respectively.

$$AML\_hops = s \times hops\_inter + s \times hops\_intra \qquad (1)$$

$$MST_{hops} = 1 \times hops_{inter} + 2 \times (s-1) \times hops\_intra \quad (2)$$

Transformating (1) and equation (2), it is easy to conclude equation (3) and equation (4) as following.

$$\Delta = MST\_hops - AML\_hops \qquad (3)$$

$$\Delta = (1-s) \times hops_{inter} + (s-2) \times hops\_intra \qquad (4)$$

Generally, hops_intra is several hops even one hop, and hops_inter is tens of hops even hundreds of hops for massive supercomputers, especially for Tianhe supercomputers. So, networking hops for hops_intra is far less thanthat of hops_inter, as following equation (5).

$$hops\_inter \ll hops\_intra \qquad (5)$$

Based on equation (4) and equation (5), we can conclude that as equation (6)

$$MST\_hops \ll AML\_hops \qquad (6)$$

Accodingly, it is easy to see that the performance of MST is much better than that of AML for transferring messages in theory.

Practically, the one-sided message in MST should follow three steps. they are (1) gathering scattered messages in comm_intra and then packing into a long message firstly, (2) secondly, forwarding the long message in omm_inter but across comm_intra, (3) finally, sattering packed message into the destination in comm_intra. Taking 4 scattered messges from comm_intra 0 to comm_intra 1, firstly,4 messages are gathered into rank_3 in comm_intra 0 and then packed into a long message; secondly, the long message would be forwarded to rank_3 comm_intra 1 in comm_inter but across comm_intra; thirdly, scattering packed message in to destination in comm_intra 1, as demonstrated in Figure 6(a). Although, there are only two steps in AML for one-sided message, in which (1) scatter messages are transferred independently in comm but across comm_intra, (2) transferred messages are forwarded in comm_intra. For example, there are 4 scatter messages fom comm_intra 0 to comm_intra 1, such as the msg 1 from rank_1 in comm_intra 0 to rank_2 in comm_intra 1, the AML should send the msg 1 from rank_1 in comm_intra 0 immediately across comm_intra to rank_1 in comm_intra 1, then forwarding the message to rank_2 in comm_intra 1, and handle remaining scattered messages in a similar flow, as as illustrated in Figure 6(b). The performance of AML would damage severely due to Scattered and separately amout of short messages transmission.

MST is not only much better than AML on One-sided Message, but also built in two-sided message, which is not supported by ML.

### 4.2 Two-sided Message

Although one-sided message could facilitate both MST and ML, there are still helpful optimizations that two-sided message would advance message transmission and win better performance than that of one-sided message in Graph500 testing on supercomputers. Unfortunately, there is no two-sided messages in AML and current version of AML support only one-sided message that cannot send a response from active message handler. However, in bottom up BFS, status of visited vertices is asked but the handler from active message could not feedback status [19], resulting in hybrid BFS failed to win expecting performance improvement. Similarly, information from wether the long edges in the buckets is crucial for switching between the Δ-stepping algorithm and the Bellman-Ford algorithm when hybridization optimization strikes to traverse from back to front. information on the long edges in the buckets could not feedbackfrom active message handler in one-sided message, which would cause failure on switching between the Δ-stepping algorithm and the Bellman-Ford algorithm and damage SSSP performane. Therefore, both BFS and SSSP are beneficial from two-sided message when testing Graph500.

In order to built two-sided message in MST, we try to xamine two-sided message using AML, as AML claim that there are several troubles we find. (1) two-sided message is advisable to divide message into segments; (2) if exchanging message is not segmented, Graph500 with AML is prone to error when inputting scale ≥ 28; (3) even in segments, Graph500 with AML is also easy to bug as fragment scale ≥ 17 . Therefore, we design and implement two-sided message mechanism based on extending one-sided message in MST, as shown in Figure 7.

As demonstrated in Figure 7, the green solid line represents forward message, while the blue dotted line denotes reverse feedback. For example, message from rank_1 forwarded by rank_3 to rank_11 with green solid line is forward sending message, while feedback from rank_11 to rank_1 is reverse feedback information that is Hrdly finished in AML as shown in Figure 6. So, based on forward sending message and reverse feedback information, two-sided message mode is built in MST.

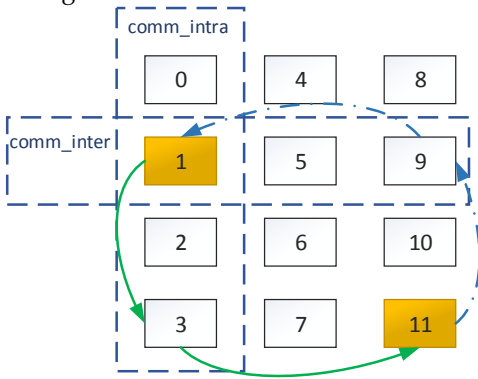

Figure 7. Two-sided message in MST

According to two-sided message built in MST, bottom up BFS optimization in Graph500 would easily get feedback status, and hybridization optimization from SSSP would feedback information easily and quickly, which would sharply boost performance when Graph500 on supercomputers, especially for Tianhe supercomputers.

## 4.3 Two-sided Message with Buffer

Furthermore, MST also provides two-sided message with buffer, which is an alternative mode for two-sided message, and would attain suprising improvement and robustness.

In both one-sided message and two-sided message detailed above, they come with a static buffer for message transition by default, as long as reaching buffer size, it would launch message transfer, which could result in frequent messages transmission and low bandwidth ulitilization.

Different from static buffer by default, we proposed a dynamic buffer for two-sided message with buffer named two-sided message with buffer, in which buffer would be dynamically expand on demand.

## 5 IMPLEMENTATION WITH EVALUATION

We propose a DAQB (Double Asynchronous Quardri Buffer) mechanism to implement MST based on MPI 3.2.1.

For DAQB, there are several rules as following:

(1). The receiving server and the sending server would receive and transmit data asynchronously using repeated non-blocking communication respectively to realize the overlapping of calculation and communication.

(2) The communication flow in MST must gather and pack scattered messages firstly in the comm_intra sub-communication domain, and then forwarding message across comm_intra but in comm_inter communication domain, finally scattering packed messages into destination in comm_intra for relieving communication traffic.

(3). Merging and sending messages according to target process with four buffers by default to lessen communication overhead.

(4). Standard MPI interface is used to transfer send and handle different types and sizes of message, so as to reduce the difficulty of programming and modification

Based on principle of DAQB, the main algorithm of implementation for MST is listed in Table 1. detsils of implementation for MST are detailed as following.

Tbale 1. The pseudo code of DAQB for implementing MST

| |
|---|
| 1.   mpi initial |
| 2.   //custom function by user for comm_inter |
| 3.   void mst_register_handler(void(*f)(int, void*, int),int n)// there are three parameters in f(int, void*, int), rank for source proess, mseesage pointer and len respectively |
| 4.   dividing communication into comm_intra and comm_inter similar with Figure 3 butcomm_inter instead of comm |
| 5.   creat buffers for receiving server and the sending server,respectively |
| 6.   voting route process in comm_intra |
| 7.   define messages msgs. |
| 8.   listening msgs from both comm_intra and comm_inter at route |
| 9.   if route.msg ->rank ∈comm_intra.route //msg from same comm_intra |

10. then
11. comm_intra.buffer[i]←msgs// according to the target process of data trans-mission
12.   if comm_intra.buffer[i] →size ≥thr  //thr is defined by ueser
13.     then
14.         gather scattered msgs to route
13.       endif
14. else
15. route.msg ->rank ∈ comm_inter.route  //msg from across comm_intra
16. call  mst_register_handler(void(*f)(int,  void*, int),int n)
17. barrier()
18. if all msgs from intra and inter to route is finised
19. then  exit
20. else   goto 8

## 5.1 Static Buffer in Comm_intra

Buffer is vital for message transimission, so static buffer is built in MST. There are sending buffer and receiving buffer for sending and receiving client, respectively.

Firstly, static buffer in comm_intra for sending sever(sendbuf_intra) is detailed following.

Besides the size of each defined by user, there are two important factors in sendbuf_intra including number of processes in comm_intra and number of handles for sending messages, as figured in Figure 7.

sendbuf_intra

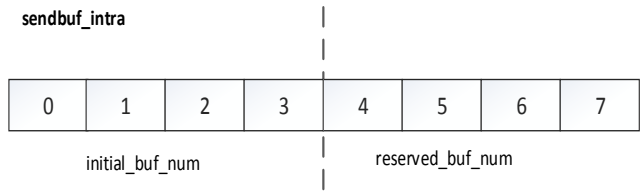

Figure 7. Static Sending Buffer in Comm_intra

It is easy to see that the number of static sending buffer(send_buf_num) in comm_intra includes intial buffer and reserved buffer, as equation (7-9).

initial_buf_num = number of processes                    (7)
$resevered\_buf\_num$ = number of handles                 (8)
send_buf_num = initial_buf_num + $resevered\_buf\_num$ (9)

in equation (7-9), initial_buf_num is not required be equal to $esevered\_buf\_num$ but they are default by four. In practice, the reserved buffer is opened togrther with the initial buffer for static buffer in comm_intra.

Similar with sendbuf_intra, static buffer in comm_intra-forreceiving sever(recvbuf_intra) is demonstrated in Figure 8.

Different from sendbuf_intra, there is no resered buffer for recvbuf_intra. And the number of recvbuf_intra(recv_buf_num) is the number of handles for receiving messages, as equation (10-11).

$recv\_buf\_num$ = initial_buf_num                  (10)
initial_buf_num = number of handles                (11)

recvbuf_intra

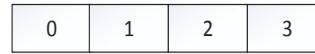

Figure 8. Static Receiving Buffer in Comm_intra

## 5.2 Active Buffer switching

Because there are initial buf and reserved buffer in sendbuf_intra, the active buffer switching policy is pecialized for sending sever.

Active Buffer isdefined as the buffer bing in use, so both initial buf and reserved buffer have chance to be active, while the other is avaible, as illustrated in Figure 9.

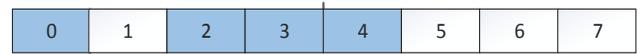

Figure 9. Active Buffer Switching in Comm_intra

For active buffer switching, the key technique is to switch active buffer between initial buffer and reserved buffer, as listed algorithm in Table 3.

| Tbale 2. Switching algorithm for active buffer |
| --- |
| //initial index array for both buffer |
| 1.   Int ini[ins], res[ind];// ins, ind are index for initial buf fer and reserved buffer |
| 2.   waiting for active_buf free |
| 3.   active_buf=initial_buf; |
| //get number of active_buf, index |
| 4.    index=ins |
| 5.   if index>=ins then |
| // 6.1 and 6.2 would  be running  in parallel |
| 6.1 sending  messages fom active buffer to other comm_intra |
|      free fom active buffer |
| 6.2   if one of handle is using then |
| // swithching active buffer, |
| 7.         active_buf= reserved buffer; |
| 8.         ins=ind; |
| 9.    goto 4 |
| 10. endif |
| 11. else  index ++ then |
| 12. goto 5 |
| 13. endif |

## 5.3 Dynamic Buffer for Two-sided Message

As detailed above, as buffers are filled up, the gathered message in comm_intra must be sent immediately even if there sre still scattered message to same comm_intra for advance bandwidth ulitilization, especially for reponse in two-sided message, which resulting in expensive message transfer across comm_intra is decided by the size and the number of buffers, not on demand. Therefore, dynamic buffer for two-sided message is proposed in MST, in which dynamic buffer expansion is built for messages delivery on demand for maximizing bandwidth ulitilization, especially for Tianhe proprietary interconnect network. In practice, two-sided message is usually used to reponse and segment entire message ino sub-segment to ease communication congestion and advance robustness, and mst_barrier_back to stop gather message on demand. The pseudo code is listed in Table 3.

Tbale 3. Dynamic Buffer for Two-sided Message

//dynamic buffer for two-sided message is used to response
ini_buf: intial buffer size imilar to pure two-sided message
cur_buf: buffer size in use
total_buf: total buffer used for response
seg_scale: optimal size of sub-segment on message, which is tunning parameters for target interconnect networking, for example seg_scale=15
bw: bandwidth ulitilization for target interconnect networking
bw_piw: optimal bandwidth ulitilization
mst_barrier_back: function to stop gather message on demand

1. total_buf= ini_buf // ini_buf is defined by user
2. dividing message with seg_scale
3. cur_buf= seg_scale
4. if (cur_buf > total_buf) then
5. total_buf= cur_buf + ini_buf
6. bw_piw= total_buf/bw
7.    if (bw_piw is ok) || (call mst_barrier_back on demand ) then
8.       call mst_barrier_back
9.    endif
10. else goto 2
11. endif

## 5.4 Experimental Setup

In order to validate proposed optimizations above for Graph500 testing on Tianhe Pre-exascale system and Tianhe AI testing platform respectively, the main Experimental Setups are listed in Table 4. From Table 4, the main differences   for testing Graph500 between Tianhe Pre-exascale system and Tianhe AI testing platform are Proprietary Interconnect and proprietary CPU. So, no special statements, sG500 mainly testing on Tianhe Pre-exascale system and *edgefactor* $=16$ .

Table 4. Configuration for Testing Graph500

| Testing system | Architecture | Parameters | Notes |
|---|---|---|---|

| | Num. of Nodes | 512 | |
| Tianhe Pre-exascale system | CPU | 2 GHz | Matrix2000+ 3x128 |
| | Cores/Node | 384 | |
| | Num. of Cores | 196608 | |
| | Memory /Node | 192 GB | |
| | Memory(GB) | 98304 | |
| | Proprietary Interconnect | | TH-Ex2 |
| | Num. of Nodes | 1024 | |
| Tianhe AI testing platform | CPU | 2.3 GHz | FT2000+ 1x64 |
| | Cores/Node | 64 | |
| | Num. of Cores | 65536 | |
| | Memory /Node | 128 | |
| | Memory(GB) | 131072 | |
| | Proprietary Interconnect | | TH-Ex2+ |

## 5.4 Communication effiiency on MST

In order to validate MST (MST without dynamic buffer for two-sided message), New-MST (MST with dynamic buffer for two-sided message), we conduct extensive performance comparison and analysis on MST, New-MST and AML.

One-sided message is the basis on communication library, and the one-sided message is built in AML and optimized in MST and New-MST. So, Communication performace on one-sided message is presented firstly as demonstrated in Table 5 and Table 6, and each testing result data below in Tables is the average of 3 exprimental results. Moreover, there is no performance difference between one-sided message and two-sided message for current New-MST.

Table 5. Communiation Time with One-sided Message on Tianhe AI Testing Platform

| Scale | Communication Library(s) | | |
|---|---|---|---|
| | AML | MST | New-MST |
| 26 | 1.063721 | 0.754269167 | 0.746565333 |
| 27 | 2.029976979 | 1.26508925 | 1.353124 |
| 28 | 4.174579 | 2.360920208 | 2.709755 |
| 29 | 9.205555938 | 4.641834917 | 4.916599333 |
| 30 | 23.81567838 | 9.634718667 | 9.539241333 |
| 31 | 79.88036304 | 26.40762088 | 20.08913933 |

As listed in Table 5, it is easy to see that communication time on both MST and New-MST is much fewer than that of AML on Tianhe AI testing platform, and as increasing scale, advantage is more and more prominent. Furthermore, the performance of New-MST is much better than that of MST as scale from 26 to 31, although there is a slight fluctuation.

Table 6. Communiation Time with One-sided Message on Tianhe Pre-exascale System

| Scale | Communication Library(s) | | |
|---|---|---|---|
| | AML | MST | New-MST |
| 26 | 1.592161021 | 1.078084771 | 1.082057333 |
| 27 | 3.307811271 | 2.080718521 | 2.093194333 |
| 28 | 7.138187167 | 4.432171979 | 4.102734667 |
| 29 | 15.55900481 | 10.70678908 | 8.34948 |
| 30 | 31.95777698 | 28.14610063 | 16.82445767 |
| 31 | 63.1955049 | 85.58677488 | out of memory |

From Table 6, we can find that as scale increasing from 26 to 30, communication time on both MST and New-MST is much fewer than that of AML on Tianhe Pre-exascale System, and the performance of New-MST is much better than that of MST as scale from 26 to 30, which is similar with behavior on Tianhe AI testing platform. While there is an abnormal decent at scale=31, and there is no result for New-MST due to out of memory. The reason for sudden drop is very ccomplicated and we would check it out when more computing node and bigger-scale testing available.

Based on One-sided Message on Tianhe AI testing platform and Tianhe Pre-exascale System, it is obvious that both MST and New-MST are much better than AML on the whole, and New_MST is superior to MST on one-sided message.

Different from One-sided Message, message segment is advisable to two-sided message. Furthermore, both MST and New-MST have two-sided message that is not well configured in AML as it claimed. In practice, AML can also avhieve two-sided message based on two-sided message but behaviors weirdly, as illustrated in Table 7 and Table 8, in which Seg_scale is the size of segment for message.

Table 7. Communiation Time on Two-sided Message with AML on Tianhe Pre-exascale System

| Scale | Seg_scale | | | |
|---|---|---|---|---|
| | 16 | 17 | 18 | >18 |
| 26 | 3.916641 | 2.609834 | - | |
| 27 | 7.154408 | 4.828078667 | - | |
| 28 | 13.82639167 | 9.627043 | 7.404867 | - |
| 29 | 25.63163333 | 17.890379 | - | |
| 30 | 48.035089 | 36.413088 | 27.567206 | |
| 31 | 91.27366425 | 65.216327 | - | |
| - | means program can not run and finish correctly | | | |

In practice, if there is no segment or even segment with Seg_scale>18 in two-sided message with AML, communication library is bound to be wrong, which resulting in that program can not run or finish correctly. Even if segment in two-sided message with AML, communication library will probably go wrong when Seg_scale=18, but when Seg_scale<18, it can run correctly on Tianhe Pre-exascale System from Table 7.

Similar with AML on Tianhe pre-exascale system, For Tianhe AI testing platform, the behavior of two-sided message with AML is just as bad, as shown in Table 8.

Table 8. Communiation Time on Two-sided Message with AML on Tianhe AI Testing Platform

| Scale | Seg_scale | | |
|---|---|---|---|
| | 16 | 17 | >=18 |
| 26 | 3.38 | 2.02 | |
| 27 | 6.09 | 3.83 | |
| 28 | 11.54 | 7.54 | |
| 29 | 21.24 | 15.79 | - |
| 30 | 41.17 | 26.75 | |
| 31 | 107.77 | 79.99 | |
| - | means program can not run and finish correctly | | |

Besides weird behavior, AML performs inefficiency compared to MST and New-MST. Due to performance of two-sided message influenced by the size of Seg_scale, so, we examine varying Seg_scale to look for an optimial Seg_scale for MST, New-MST and AML according to target proprietary interconnect networking built in Tianhe supercomputers, as illustrated in Figure 10 and Table 9.

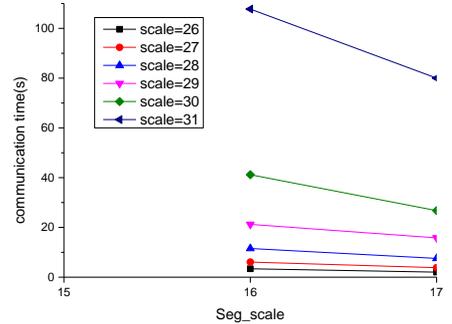

(1). Looking for optimial Seg_scale in AML

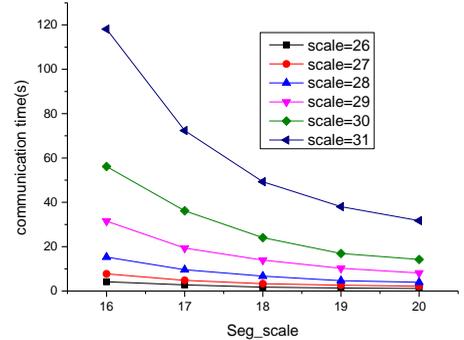

(2). Looking for optimial Seg_scale in MST

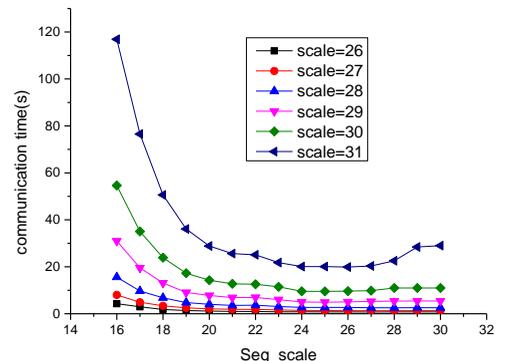

(3). Looking for optimial Seg_scale in New-MST

Figure 10. Looking for optimial Seg_scale on on Tianhe AI Testing Platform

From Figure 10(1), Figure 10(2), and Figure 10(3), it is easy to see that the optimal optimial Seg_scale is Seg_scale = 17 , Seg_scale = 20 and Seg_scale = 26 for AML, MST and New-MST, respectively.

Similarly, the optimal Seg_scale for AML, MST and New-MST on Tianhe Pre-exascale System are summarized in Table 9.

Table 9. Optimial Seg_scale on on Tianhe Pre-exascale System

|                   | AML | MST | New-MST |
|-------------------|-----|-----|---------|
| Optimial Seg_scale | 17  | 19  | 26      |

Comparing Figure 10 and Table 9, we can find that the optimal Seg_scale on both Tianhe AI testing platform and Tianhe pre-exascale system are very close, even the same, the reason is that proprietary interconnect in both Tianhe AI testing platform and Tianhe pre-exascale system are from the same origine fast interconnect networking.

Based on the optimial Seg_scale, performane on two-sided message from AML, MST and New-MST is detailed as following Table 10 and Table 11.

Table 9. Communiation Time with Two--sided Message on Tianhe AI Testing Platform

| Scale | Communication Library(s) | | |
|-------|-------------|-------------|-------------|
|       | AML         | MST         | New-MST     |
| 26    | 2.0203      | 1.149328333 | 0.746565333 |
| 27    | 3.8278265   | 2.166702333 | 1.353124    |
| 28    | 7.544396333 | 3.950838667 | 2.709755    |
| 29    | 15.79315733 | 8.121443    | 4.916599333 |
| 30    | 26.75193267 | 14.227874   | 9.539241333 |
| 31    | 79.98521867 | 31.72993933 | 20.08913933 |

As detailed in Table 9, communiation time on both MST and New-MST is fewer and fewer than that of AML and as scale increasing from 26 to 31, and as increasing scale, New-MST is better and better than MST on Tianhe AI Testing Platform.

Table 10. Communiation Time with Two--sided Message on Tianhe Pre-exascale System

| Scale | Communication Library(s) | | |
|-------|-------------|-------------|---------------|
|       | AML         | MST         | New-MST       |
| 26    | 2.609834    | 1.680606333 | 1.082057333   |
| 27    | 4.828078667 | 3.098419333 | 2.093194333   |
| 28    | 9.627043    | 5.841738333 | 4.102734667   |
| 29    | 17.890379   | 11.48876267 | 8.34948       |
| 30    | 36.413088   | 21.95226167 | 16.82445767   |
| 31    | 65.216327   | 44.232449   | out of memory |

Similar with two--sided message on Tianhe AI testing platform, both MST and New-MST are much better than AML, and New-MST is superior to MST on Tianhe pre-exascale system as scale increasing from 26 to 30 as demonstrated in Table 10.

Summarily, communication efficiency is defined as equation (12), in which Comm_Efficiency represents the efficiency of communication, comm_Volume is the total communication message volume, and Comm_Time is the spending time for message transfer.

Comm_Efficiency = comm_Volume/ Comm_Time  (12)

So, the communication efficiency comparisons on AML, MST and New-MST on Tianhe Pre-exascale System are demonstrated in Figure 11-1, Figure 11-2 and Figure 11-3.

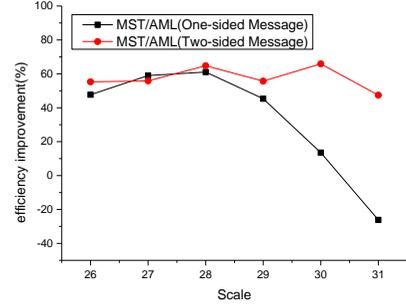

Figure 11-1. efficiency improvement on MST/AML for Tianhe Pre-exascale System

As demonstrated in Figure 11-1, MST is rather better than AML, especially for two-sided message with up to 60% communication efficiency improvement. For one-sided message, MST is superior to MST when scale is nomore than 30, while the efficiency improvement is decending from scale=28, and MST does not keep the edge to scale=31. The main reason is that immediately forwarding is not conductive to the full ulitilization of networking bandwidth if there is no dynamic buffer configured in two-sided message. Hence, New-MST with dynamic buffer is proposed to advance bandwith ulitilization for two-sided message, as illustrated in Figure 11-2.

Based on MSTand further optimitization, New-MST has overwhelming adtantages over AML as shown in Figure 11-2, in which not only one-sided message but also two-sided message is more and more better than AML as increasing scale.

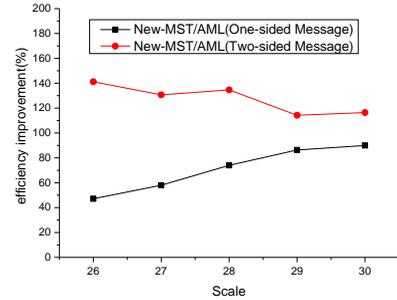

Figure 11-2. efficiency improvement on New-MST/AML for Tianhe Pre-exascale System

Moreover, performance comparison between New-MST and MST is also shown in Figure 11-3.

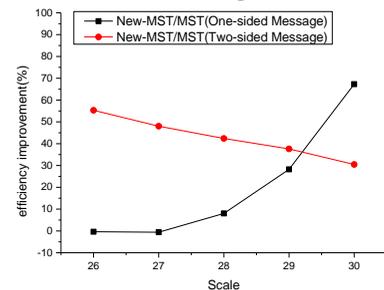

Figure 11-3. efficiency improvement on New-MST/MST for Tianhe Pre-exascale System

Figure 11. efficiency improvement on Tianhe Pre-exascale System

From Figure 11-3, it is easy to see that New-MST is much better than MST as scale from 26 30 for one-sided message and two-sided message. According to Figure 11-1, Figure 11-2 and Figure 11-3, we find that there is no data when running New-MST at scale=31 because of out of memory.

Furthermore, the communication efficiency comparisons are also validated on Tianhe AI Testing Platform, as demonstrated in Figure 12.

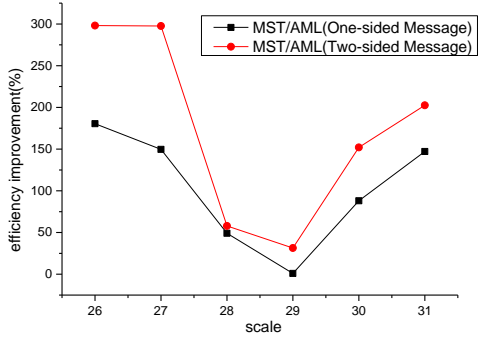

(1). efficiency improvement on MST/AML for Tianhe AI Testing Platform

From Figure 12(1), it can find that MST performs rather better than AML on both one-sided message and two-sided message on Tianhe AI Testing Platform.

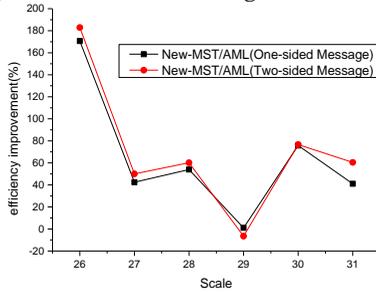

(2). efficiency improvement on New-MST/AML for Tianhe AI Testing Platform

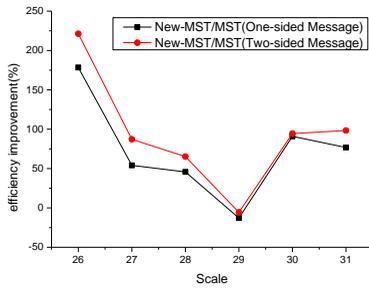

(3). efficiency improvement on New-MST/MST for Tianhe AI Testing Platform

Figure 12. efficiency improvement on Tianhe Pre-exascale System

According to Figure 11 and Figure 12, it can conclude that not only MST but also New-MST are rather better than AML on both Tianhe superompters, moreover, both MST and New-MST behavior better on Tianhe AI Testing Platform than that of Tianhe Pre-exascale System. The main reason is the difference on proprietary interconnect built in Tianhe superompters.

## 5.5 Graph500 with MST

In oerder to validate pratical effect on large-scale application, we test Graph500 based on BFS with MST on Tianhe supercomputers, as demonstrated in Figure 13.

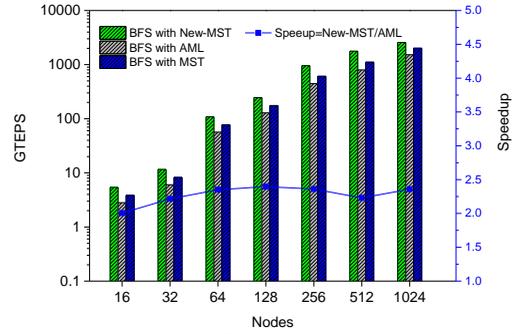

(1). Graph500 with BFS testing on Tianhe AI Testing Platform

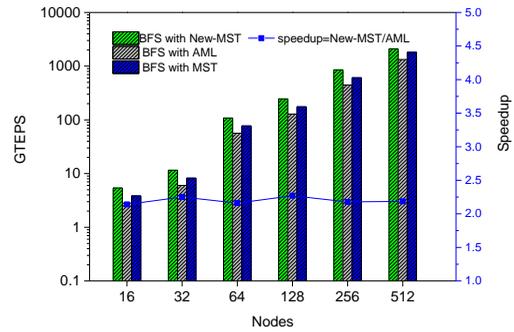

(2). Graph500 with BFS testing on Tianhe Pre-exascale System

Figure 13. Graph500 with BFS testing on Tianhe supercomputers

Obviously, Graph500 based on BFS with New-MST would win the best performance on Tianhe supercomputers, and Graph500 based on BFS with MST could attain higher GTEPS than that of Graph500 based on BFS with AML, and the maximal sppedups on New-MST/ AML are both close to 2.5 times for Tianhe supercomputers from Figure 13.

Extensively, Graph500 based on SSSP as an alternative bencmarch is also testing with MST, as illustrated in Figure 14.

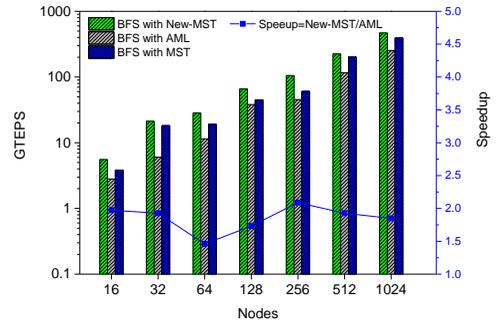

(1). Graph500 with SSSP testing on Tianhe AI Testing Platform

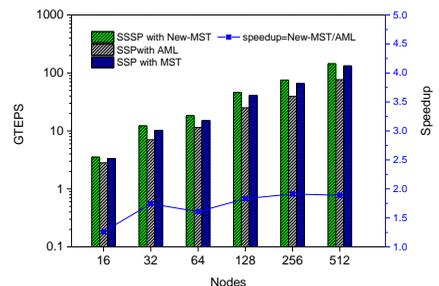

(2). Graph500 with SSSP testing on Tianhe Pre-exascale System

Figure 14. Graph500 with SSSP testing on Tianhe supercomputers

Similar with Graph500 with BFS, Graph500 based on SSSP with New-MST is much better than that of MST and AML, and Graph500 based on SSSP with MST is better than that of AML, Differently, the maximal sppedups on New-MST/ AML is slightly smaller than that of Graph500 based on SSSP for Tianhe supercomputers. That is because program features and operators vary between BFS and SSSP.

## 6 CONCLUSION AND DISCUSSION

The communication efficiency is critical to lasrge-scale graph on supercomputers, so we present MST, an efficient message transfer library for graph computation like Graph500 on supercomputers, especially for Tianhe supercomputers.

Our MST not only includes one-sided message, but also extend two-sided messages. Furthermore, we propose New-MST with dynamic buffer for two-sided message to further advance communication efficiency. For both MST and New-MST, message segment is advisable to two-sided messages, and the size of segment should be tunning according target interconnect network. Experimental results show that both MST and New-MST are much better than AML, and the maximal sppedups on Graph 500 with New-MST/ Graph 500 with AML are close to 2.5 times and close to 2.0 times for Tianhe AI Testing Platform and Tianhe Pre-exascale System, respectively.

For the incoming Tianhe exascale system, we will be facing many new challenges. the MST scalability and further optimization on New-MST as well as extending more large-scale applications using MST by planned open source MST are our new directions.

## REFERENCES

[1]  Top500 : http://www.top500.org/.
[2]  Ruibo Wang, Kai Lu, Juan Chen, Wenzhe Zhang, Jinwen Li, Yuan Yuan, Pingjing Lu, Libo Huang, Shengguo Li, and Xiaokang Fan." Brief Introduction of TianHe Exascale Prototype System". TSINGHUA SCIENCE AND TECHNOL-OGY ,2021,26(3): 361–369.
[3]  Graph500 : http://www.graph500.org/.
[4]  Koji Ueno and Toyotaro Suzumura "Highly Scalable Graph Search for the Graph500 Benchmark". In Proceedings of the 21st International ACM Symposium on High-Performance Parallel and Distributed Computing2012.
[5]  Pablo Fuentes, Jos´e Luis Bosque and Ram´on Beivide. "Characterizing the Communication Demands of the Graph500 Benchmark on a Commodity Cluster". In Proceedings of the IEEE/ACM International Symposium on Big Data Computing,2014.
[6]  Toyotaro Suzumura, Koji Ueno, Hitoshi Sato, Katsuki Fujisawa and Satoshi Matsuoka. "Performance Characteristics of Graph500 on Large-Scale Distributed Environment". In Proceedings of the IEEE International Symposium on Workload Characterization (IISWC)，2011.
[7]  Masahiro Nakao, Koji Ueno, Katsuki Fujisawa, Yuetsu Kodama and Mitsuhisa Sato. "Performance Evaluation of Supercomputer Fugaku using Breadth-First Search Benchmark in Graph500". In Proceedings of the IEEE International Conference on Cluster

Computing (CLUSTER),2020.
[8]  Koji Ueno and Toyotaro Suzumura. "2D Partitioning Based Graph Search for the Graph500 Benchmark". In Proceedings of the IEEE 26th International Parallel and Distributed Processing Symposium Workshops & PhD Forum,2012.
[9]  M. Blocksome, C. Archer, T. Inglett, et.al. "Design and Implementation of a One-Sided Communication Interface for the IBM eServer Blue Gene Supercomputer". In Proceedings of the 2006 ACM/IEEE SC|06 Conference (SC'06),2006.
[10]  Ahmad Faraj, Sameer Kumar, Brian Smith, et.al. "MPI Collective Communications on The Blue Gene/P Supercomputer: Algorithms and Optimizations". In Proceedings of the 17th IEEE Symposium on High Performance Interconnects,2009.
[11]  Hao Yu, I-Hsin Chung, Jose Moreira." Topology Mapping for Blue Gene/L Supercomputer". In Proceedings of the 2006 ACM/IEEE SC|06 Conference (SC'06).
[12]  Sameer Kumar, Amith R. Mamidala, Daniel A. Faraj, et.al. "PAMI: A Parallel Active Message Interface for the Blue Gene/Q Supercomputer". In proceedings of the IEEE 26th International Parallel and Distributed Processing Symposium,2012.
[13]  Gautam Shah, Jarek Nieplocha, Jamshed Mirza, et.al. "Performance and Experience with LAPI – a New High-Performance Communication Library for the IBM RS/6000 SP". In proceedings of the First Merged International Parallel Processing Symposium and Symposium on Parallel and Distributed Processing.1998.
[14]  Naoyuki Shida, shinji Sumimoto, Atsuya Uno. "MPI Library and Low-Levemcommunication on the K computer". FUJISTU sci.tech.J., vol,48, No.3, 2012.
[15]  Masahiro Nakao, Koji Ueno, Katsuki Fujisawa, Yuetsu Ko-dama and Mitsuhisa Sato. "Performance Evaluation of Su-percomputer Fugaku using Breadth-First Search Benchmark in Graph500". In Proceedings of the IEEE International Con-ference on Cluster Computing (CLUSTER),2020.
[16]  Mingzhe Li, Xiaoyi Lu, Sreeram Potluri, et.al. "Scalable Graph500 Design with MPI-3 RMA". In proceedings of the IEEE International Conference on Cluster Computing (CLUS-TER),2014.
[17]  Shang Li, Po-Chun Huang and Bruce Jacob." Exascale Inter-connect Topology Characterization and Parameter Explora-tion". In proceedings of the 20th IEEE International Confer-ence on High Performance Computing and Communica-tions,2018.
[18]  Yi Zhu, Michael Taylor, Scott B. Baden and Chung-Kuan Cheng." Advancing supercomputer performance through in-interconnection topology synthesis". In proceedings of the IEEE/ACM International Conference on Computer-aided Design,2018.
[19]  Scott Beamer, Krste Asanović and David Patterson. "Direction-optimizing breadth-first search". In Proceedings of the International Conference on High Performance Computing, Networking, Storage and Analysis,2012.
[20]  Venkatesan T. Chakaravarthy, Fabio Checconi, Prakash Murali, et.al. "Scalable Single Source Shortest Path Algorithms for Massively Parallel Systems". IEEE Transactions On Parallel and Distributed Systems, Vol. 28, No. 7, July 2017.